\documentclass[twocolumn,floatfix]{revtex4}
	\usepackage{amssymb}
	\usepackage{epsfig}
	\usepackage{amsmath}
	\usepackage{graphicx}
	
\begin{document}

\newcommand{\comment}[1]{}

\newcommand{\traceroute}{traceroute}
\newcommand{\traceroutes}{traceroutes}

\newcommand{\eps}{\epsilon}
\newcommand{\e}{{\rm e}}
\newcommand{\dm}{{\rm d}m}
\newcommand{\dx}{{\rm d}x}
\newcommand{\ds}{{\rm d}s}
\newcommand{\dt}{{\rm d}t}
\newcommand{\du}{{\rm d}u}
\newcommand{\ex}[1]{{\rm E}[#1]}
\newcommand{\poi}{{\rm Poisson}}

	\title{Traceroute sampling makes random graphs appear to have power law degree distributions}
	\author{Aaron Clauset$^*$ and Cristopher Moore$^{\dagger,*}$}
	\affiliation{$^*$Computer Science Department and $^\dagger$Department of Physics and Astronomy, University of New Mexico, Albuquerque NM 87131 \\ 
	{\tt \{aaron,moore\}@cs.unm.edu}}
	\date{\today}

\begin{abstract}
The topology of the Internet has typically been measured by sampling {\em traceroutes}, which are roughly shortest paths from sources to destinations.  The resulting measurements have been used to infer that the Internet's degree distribution is scale-free; however, many of these measurements have relied on sampling traceroutes from a small number of sources.  It was recently argued that sampling in this way can introduce a fundamental bias in the degree distribution, for instance, causing random (Erd\H{o}s-R\'enyi) graphs to appear to have power law degree distributions.  We explain this phenomenon analytically using differential equations to model the growth of a breadth-first tree in a random graph $G(n,p=c/n)$ of average degree $c$, and show that sampling from a single source gives an apparent power law degree distribution $P(k) \sim 1/k$ for $k \lesssim c$.
\comment{
Sampling the Internet topology via traceroute methods sourced at a single or small number of hosts is akin to using a spanning tree on the Internet to infer topological features. Specifically, such methods have been used to infer that the Internet's router degree distribution is scale free. It was recently argued that such a sampling method can introduce fundamental bias in the resulting topological inferences, and that these biases allow a sparse Erd\H{o}s-R\'enyi graph, with Poisson degree distribution, to appear to have a powerlaw degree distribution. We explain analytically, using a differential equations model of the growth of a random spanning tree, how this process produces such biased results.
}
\end{abstract}
\maketitle

\section{Introduction}
The Internet and the networks it facilitates --- including the Web and email networks --- are the largest artificial complex networks in existence, and understanding their structural and dynamic properties is important if we wish to understand social and technological networks in general. Moreover, efforts to design novel dynamic protocols for communication and fault tolerance
are well served by knowing these properties.

One structural property of particular interest is the degree distribution at the router level of the Internet. This distribution has been inferred~\cite{Govindan, Faloutsos, Rocketfuel, CAIDA, Oregon, LookingGlass, Opte} both by
sampling {\em \traceroutes}, i.e., the paths chosen by Internet routers, which approximate shortest paths in the network, and by taking ``snapshots'' of BGP (Border Gateway Protocol) routing tables~\cite{BGP}.  These methods have been criticized as being noisy and imperfect~\cite{Amini, Chen}.  However, Lakhina et al.~\cite{Lakhina} recently argued that such methods have a more fundamental flaw. Due to the fact that these methods use a small number of sources for the inference, they were able to show that only a small fraction of the edges of a graph are ``visible'' in such a sample.  Moreover, the set of visible edges is biased in such a way that Erd\H{o}s-R\'enyi random graphs~\cite{gnp}, whose underlying distribution is Poisson, will appear to have a power law degree distribution.  While no one would argue that the Internet is a purely random graph, this certainly calls into question the standard measurements of power law or ``scale-free'' degree distributions on the Internet, and reopens the problem of characterizing Internet topology. 

In this paper we explain this bias phenomenon analytically.  Specifically, by modeling the growth of a breadth-first spanning tree with differential equations, we show that sampling shortest paths from a single source in an Erd\H{o}s-R\'enyi random graph gives rise to a power law degree distribution of the form $P(k) \sim 1/k$, up to a cutoff $k \sim c$ where $c$ is the average degree of the underlying graph.  While sampling traceroutes from a single source is rather limited, Barford~\cite{Barford} provides empirical evidence that, on the Internet, merging shortest paths from several sources leads to only marginally improved surveys of Internet topology.  In the Conclusions we discuss generalizing our approach to sampling from multiple sources.

Finally, even if the Internet has a power-law degree distribution, the exponent may be rather different from the one observed in \traceroute\ samples.  In future work we plan to extend our approach to graphs with arbitrary degree distributions, to study the relationship between the observed exponent and the underlying one.

\section{Internet Spanning Trees}

Most mapping projects have inferred the Internet's degree distribution by implicitly building a map of the network from the union of a large number of \traceroutes\ from a single source, or from a small number of sources.  Assuming that \traceroutes\ are shortest paths, sampling from a single source is equivalent to building a spanning tree.  We therefore model this sampling method by modeling the growth of a spanning tree on a graph.

There are several ways one might build a spanning tree.  We will consider a family of methods, in which at each step, every \mbox{vertex} in the graph is labeled {\em reached}, {\em pending}, or \mbox{{\em unknown}}.  The pending vertices are the leaves of the current tree, the reached vertices are those in its interior, and the unknown vertices are those not yet connected.  To model \traceroutes\ from a single source, we initialize the process by labeling the source vertex pending, and all other vertices unknown.  

Then the growth of the spanning tree is given by the following pseudocode.
\begin{quote}
\begin{tabbing}
{\tt while}\=\ there are pending vertices: \\
\>choose a pending vertex $v$ \\
\>label $v$ reached \\
\>for \=every unknown neighbor $u$ of $v$, \\
\> \> label $u$ pending.
\end{tabbing}
\end{quote}

The type of spanning tree is determined by how we choose which pending vertex $v$ we will use to extend the tree.  To model shortest paths, we store the pending vertices in a queue, take them in FIFO (first-in, first-out) order, and build a breadth-first tree; if we like we can break ties randomly between vertices of the same age in the queue, which is equivalent to adding a small noise term to the length of each edge as in~\cite{Lakhina}.  If we store the pending vertices on a stack and take them in LIFO (last-in, first-out) order, we build a depth-first tree.  Finally, we can choose from among the pending vertices uniformly at random, giving a ``random-first'' tree.

\begin{figure} [htbp]
\begin{center}
\includegraphics[scale=0.45]{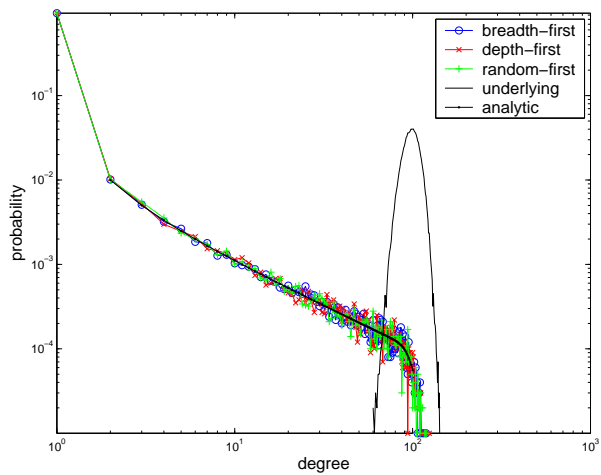}
\caption{A log-log plot of sampled degree distributions from breadth-first, depth-first and random-first spanning trees on a random graph of size $n=10^5$ and average degree $c=100$, 
and our analytic results (the black dots).  The agreement between the three types of trees and our analytic results is extremely good.  For comparison, the black line shows the Poisson degree distribution of the underlying graph.  Note the power law behavior of the apparent degree distribution for $k$, which extends up to a cutoff at $k \sim c$. }
\label{fig:graph}
\end{center}
\end{figure}

Surprisingly, while these three processes build different trees, and traverse them in different orders, we will see in the next section that they all yield the same degree distribution when $n$ is large.
To illustrate this, Fig.~\ref{fig:graph} shows empirical degree distributions of breadth-first, depth-first, and random-first spanning trees for a random graph $G(n,p=c/n)$ where $n=10^{5}$ and $c=100$.  The three degree distributions are indistinguishable; further, they are well-matched by the analytic results given in the next section, and obey a power law with exponent $-1$ for degrees less than $c$.  For comparison, we also show the Poisson degree distribution of the underlying graph.


\section{Analytic Results}

In this section we show analytically that building spanning trees in Erd\H{o}s-R\'enyi random graphs $G(n,p=c/n)$, using any of the processes described above, gives rise to an apparent power law degree distribution $P(k) \sim 1/k$ for $k \lesssim c$.  We focus here on the case where the average degree $c$ is large, but constant with respect to $n$; we believe our results also hold if $c$ is a moderately growing function of $n$, such as $\log n$ or $n^\eps$ for small $\eps$, but it seems more difficult to make our analysis rigorous in that case.


We will model the progress of the {\tt while} loop described above.  Let $S(T)$ and $U(T)$ denote the number of pending and unknown vertices at step $T$ respectively.  The expected changes in these variables at each step are (where $\ex{\cdot}$ denotes the expectation)
\begin{eqnarray}
\ex{U(T+1)-U(T)} & = & - p U(T) \nonumber \\
\ex{S(T+1)-S(T)} & = & p U(T) - 1 \label{eq:diff}
\end{eqnarray}
Here the $p U(T)$ terms come from the fact that a given unknown vertex $u$ is connected to the chosen pending vertex $v$ with probability $p$, in which case we change its label from unknown to pending; the $-1$ term comes from the fact that we also change $v$'s label from pending to reached.  Moreover, these equations apply no matter how we choose $v$; whether $v$ is the ``oldest'' vertex (breadth-first), the ``youngest'' one (depth-first), or a random one (random-first). By the principle of deferred decisions, the events that $v$ is connected to each unknown vertex $u$ are independent and occur with probability $p$.  Our experiments do indeed show that these three processes result in the same degree distribution.

Writing $t=T/n$, $s(t) = S(tn)/n$ and $u(tn)=U(t)/n$, the difference equations~\eqref{eq:diff} become the following system of differential equations,
\begin{eqnarray}
\frac{\du}{\dt} & = & -c u \nonumber \\
\frac{\ds}{\dt} & = & c u - 1 
\label{eq:sys}
\end{eqnarray}
With the initial conditions $u(0) = 1$ and $s(0) = 0$, the solution to~\eqref{eq:sys} is
\begin{equation}
\label{eq:sol}
u(t) = \e^{-ct} , \quad s(t) = 1 - t - \e^{-ct} \enspace .
\end{equation}
The algorithm ends at the smallest positive root $t_f$ of $s(t) = 0$; using Lambert's function $W$, defined as $W(x) = y$ where $y \e^y = x$, we can write
\begin{equation}
\label{eq:t0}
 t_f = 1 + \frac{1}{c} W(-c\e^{-c}) \enspace .
\end{equation}
Note that $t_f$ is the fraction of vertices which are reached at the end of the process, and this is simply the size of the giant component of $G(n,c/n)$.

Now, we wish to calculate the degree distribution $P(k)$ of this tree.  The degree of each vertex $v$ is the number of its previously unknown neighbors, plus one for the edge by which it became attached (except for the root).  Now, if $v$ is chosen at time $t$, in the limit $n \to \infty$ the probability it has $k$ unknown neighbors is given by the Poisson distribution with mean $m = cu(t)$,
\[ \poi(m,k) = \frac{\e^{-m} m^k}{k!} \enspace . \]
Averaging over all the vertices in the tree 
gives
\[ P(k+1) = \frac{1}{t_f} \int_0^{t_f} \dt \,\poi(cu(t),k) \enspace . \]
It is helpful to change the variable of integration to $m$.  Since $m = c \e^{-ct}$ we have $\dm = -cm \,\dt$, and 
\begin{eqnarray}
 P(k+1) & = & \frac{1}{t_f} \int_{c(1-t_f)}^c \dm \,\frac{\poi(m,k)}{c m} \nonumber \\
 & \approx & \int_{c \e^{-c}}^c \dm \,\frac{\poi(m,k)}{cm} \nonumber \\
 & = & \frac{1}{c k!} \int_{c \e^{-c}}^c \dm \,\e^{-m} m^{k-1} \enspace .
 \label{eq:p1}
\label{eq:int}
\end{eqnarray}
Here in the second line we use the fact that $t_f \approx 1 - \e^{-c}$ when $c$ is large (i.e., the giant component encompasses almost all of the graph).

The integral in~\eqref{eq:int} is given by the difference between two incomplete Gamma functions.  However, since the integrand is peaked at $m=k-1$ and falls off exponentially for larger $m$, for $k \lesssim c$ it coincides almost exactly with the full Gamma function $\Gamma(k)$.  Specifically, for any $c > 0$ we have
\[ \int_0^{c \e^{-c}} \dm \,\e^{-m} m^{k-1} < c \e^{-c} \]
and, if $k - 1 = c (1-\eps)$ for $\eps > 0$, then
\begin{eqnarray*}
 \int_c^\infty \dm \,\e^{-m} m^{k-1} 
 & = & \e^{-c} c^{k-1} \int_0^\infty \dx \,\e^{-x} (1+x/c)^{k-1} \\
 & < & \e^{-c} c^{k-1} \int_0^\infty \dx \,\e^{-x} \e^{x(k-1)/c} \\
 & = & \frac{\e^{-c} c^{k-1}}{\eps} 
 < \frac{\e^{-(k-1)} (k-1)^{k-1}}{\eps} \\
& \approx & \frac{\Gamma(k)}{\eps \sqrt{2 \pi (k-1)}}
 \end{eqnarray*}
\comment{
and, if $k - 1 = c - \Delta$ for $\Delta > 0$, then
\begin{eqnarray*}
 \int_c^\infty \dm \,\e^{-m} m^{k-1} 
 & = & \e^{-c} c^{k-1} \int_0^\infty \dx \,\e^{-x} (1+x/c)^{k-1} \\
 & < & \e^{-c} c^{k-1} \int_0^\infty \dx \,\e^{-x} \e^{x(k-1)/c} \\
 & = & \frac{\e^{-c} c^k}{\Delta} \\
 & < & \frac{c}{\Delta} \,\e^{k-1} (k-1)^{k-1} \\
 & \approx & \frac{c}{\Delta} \frac{\Gamma(k)}{\sqrt{2 \pi k}} \enspace .
\end{eqnarray*}
}
This is $o(\Gamma(k))$ if $\eps \gtrsim 1/\sqrt{k}$, i.e., if $k < c - c^\alpha$ for some $\alpha > 1/2$.  In that case we have
\begin{equation}
P(k+1) = (1-o(1)) \frac{\Gamma(k)}{c k!} 
\sim \frac{1}{ck}
\end{equation}
giving a power law of exponent $-1$ up to $k \sim c$.

Although we omit some technical details, this derivation can be made mathematically rigorous using results of Wormald~\cite{Wormald}, who showed that under fairly generic conditions, the state of discrete stochastic processes like this one is well-modeled by the corresponding rescaled differential equations.  Specifically it can be shown that if we condition on the initial source vertex being in the giant component, then with high probability, for all $t$ such that $0 < t < t_f$, $U(tn) = u(t)n + o(n)$ and $S(tn)=s(t)n+o(n)$.  It follows that with high probability our calculations give the correct degree distribution of the spanning tree within $o(1)$.


\section{Discussion and Conclusions}

Lakhina et al.~\cite{Lakhina}, argued that sampling \traceroutes\ from a small number of sources profoundly underestimates the degrees of vertices far from the source, and that this effect can cause graphs to appear to have a power law degree distribution even when their underlying distribution is Poisson.  In this work, we modeled this sampling process as the construction of a spanning tree, and showed analytically that for sparse random graphs of large average degree, the apparent degree distribution does indeed obey a power law of the form $P(k)\sim k^{-1}$ for $k$ below the average degree.  This illustrates the danger of concluding the existence of a power-law from data over too small a range of degrees, and, more specifically, the danger of sampling \traceroutes\ from just a few sources.  While our analytic results hold for \traceroutes\ from a single source, we conjecture that if we use any constant number of sources and take the union of the resultant spanning trees, the observed distribution will remain subject to the same sampling bias, and one will again observe a power-law degree distribution.  

Certainly this exponent $-1$ is not the one observed for the Internet.  For instance, Faloutsos, Faloutsos and Faloutsos~\cite{Faloutsos} observed degree distributions $P(k) \sim k^{-2.5}$ at the router level, and $k^{-2.2}$ for out-degrees at the inter-domain (BGP) level.  While the real degree distribution of the Internet may be a power law, one possibility is that the true exponent is rather different from the observed one; this possibility was recently explored by Peterman and de los Rios~\cite{paolo} (they also study another mechanism for observing apparent power laws in the degree distribution, which samples the original graph via a probabilistic pruning strategy and gives an observed exponent of $-2$ for random graphs). In future work we will extend our differential equation model to random graphs with power law degree distributions to explore this possibility.  

\comment{Our current result further illustrates the danger of concluding the existence of a power-law form in the degree distribution from data over too small a range of degrees; indeed, the data used to infer the power-law form of the distribution for Internet routers~\cite{Faloutsos} does not span a large enough range to be beyond doubt.}

\comment{
Although our analytic model of the routing paths discovered by sampling \traceroutes\ illustrates one possible mechanism for the observed power law degree distribution in~\cite{Faloutsos}, the model may not fully explain its observation. The interaction between the routing decisions (such as those made by the border-gateway protocol) and the actual link-level topology of the Internet may give rise to an {\em effective} degree distribution which has a powerlaw form, comprised of those edges which are actually used by Internet applications as opposed to one which samples all existing edges. This possibility implies that even if the real degree distribution of the Internet is something very different from a power law, it may not matter as these extra links are not utilized in normal routing decisions. Another difference of unknown, but perhaps small, significance is that on the Internet, the routing algorithms are intended to optimize for end-to-end traffic flow, i.e., routers in the ``interior'' of the Internet such as those on the backbone are not themselves destinations or sources. Unlike the empirical studies cited earlier, we selected sources uniformly at random for sourcing the topological mapping.
}

Although our analytic model of spanning trees illustrates one possible mechanism for apparent power-law degree distributions, another possibility is that decisions (such as those made by the border-gateway protocol) made by actual Internet routers interact in complex ways with the the underlying link-level topology.  It may well be that routers only use a small fraction of the edges in the network, and that the edges they actually use give rise to an {\em effective} degree distribution with roughly a power-law form.  This possibility implies that even if the real degree distribution of the Internet is something very different from a power law, it may not matter as these extra links are not utilized in normal routing decisions. 

Another difference of unknown, but perhaps small, significance is that on the Internet, routing algorithms are intended to optimize for end-to-end traffic flow, and routers in the ``interior'' of the Internet such as those on the backbone are not themselves destinations or sources. Unlike the empirical studies cited earlier, we selected sources uniformly at random for our spanning trees.

The story of the topology of the Internet is far from over and will undoubtedly remain a topic of great interest for many years to come. However, knowing decisively that current methods are fundamentally biased will serve to push the state-of-the-art forward toward more robust methods of characterization. We see the following three lines of inquiry as enticing:

First, having analytically shown that a Poisson degree distributions can lead to apparent power laws, we naturally wish to generalize our approach to random graphs with arbitrary degree distributions, characterize the family of distributions which generate apparent power laws, and understand the relationship between observed and underlying exponents for power law distributions.  Having this generalization may allow us to make firm and useful claims about the topology of the Internet.  

Secondly, in what manner is the \traceroute-sampled degree distribution of a random graph dependent on the length of the paths taken between a source and destination?  If we simulate a ``crawl'' by using  random walks rather than short paths, how would the observed degree distribution change? 

Finally, if we use build spanning trees from $m$ sources and take their unions, how does the observed degree distribution vary with $m$?  If we are correct that any constant $m$ leads to the same kind of bias, how does $m$ need to grow with the network to obtain an accurate sample?

\begin{acknowledgments}
The authors are grateful to David Kempe, Mark Newman, Michel Morvan, Paolo De Los Rios, Mark Crovella and Tracy Conrad for helpful conversations, and to American Airlines for providing a conducive working environment.  This work was funded by NSF grant PHY-0200909.
\end{acknowledgments}



\begin{thebibliography}{99}
\bibitem{BGP}
The Border Gateway Protocol (BGP) is a distributed method by which routers which bridge two distinct networks construct forwarding tables based on the IP addresses which are ``reachable'' from the current router via ``advertised paths'' of other BGP-routers. Such routers decide which packets cross to and from the network it guards, and may also serve to forward packets between networks.
\end{thebibliography}
\end{document}